\documentclass[prl,final,twocolumn,showpacs,floatfix,%
superscriptaddress,nofootinbib]{revtex4}

\usepackage[utf8]{inputenc}
\usepackage{amsfonts}
\usepackage{amsmath}
\usepackage{amssymb}
\usepackage{graphicx}
\usepackage{nicefrac}

\providecommand{\U}[1]{\protect\rule{.1in}{.1in}}

\begin{document}

\title{Topological phases for bound states moving in a finite volume}

\author{Shahin Bour}
\affiliation{Helmholtz-Institut für Strahlen- und Kernphysik (Theorie)\\
and Bethe Center for Theoretical Physics, Universität Bonn, 53115 Bonn,
Germany}

\author{Sebastian König}
\affiliation{Helmholtz-Institut für Strahlen- und Kernphysik (Theorie)\\
and Bethe Center for Theoretical Physics, Universität Bonn, 53115 Bonn,
Germany}

\author{Dean Lee}
\affiliation{Department of Physics, North Carolina State University, Raleigh,
NC 27695, USA}

\author{H.-W. Hammer}
\affiliation{Helmholtz-Institut für Strahlen- und Kernphysik (Theorie)\\
and Bethe Center for Theoretical Physics, Universität Bonn, 53115 Bonn,
Germany}

\author{Ulf-G.~Meißner}
\affiliation{Helmholtz-Institut für Strahlen- und Kernphysik (Theorie)\\
and Bethe Center for Theoretical Physics, Universität Bonn, 53115 Bonn,
Germany}
\affiliation{Institut für Kernphysik, Institute for Advanced Simulation and
Jülich Center for Hadron Physics, Forschungszentrum Jülich, D-52425
Jülich, Germany}

\begin{abstract}
We show that bound states moving in a finite periodic volume have an energy
correction which is topological in origin and universal in character.  The
topological volume corrections contain information about the number and mass
of the constituents of the bound states.  These results have broad 
applications to lattice calculations involving nucleons, nuclei, hadronic 
molecules, and cold atoms.  We illustrate and verify the analytical results with
several numerical lattice calculations.
\end{abstract}

\maketitle

Over two decades ago Lüscher derived a relation connecting the energy levels of
an interacting two-body system in a periodic cube to physical scattering phase
shifts~\cite{Luscher:1986pf,Luscher:1991ux}. This finite-volume technique has
become a standard tool in lattice quantum chromodynamics~\cite{Beane:2003da,
Li:2007ey,Gockeler:2008kc,Feng:2009ij,Beane:2010hg,Dudek:2010ew} and in lattice
effective field theory for nucleons and cold atomic systems~\cite{Abe:2007ff,
Lee:2008fa,Epelbaum:2009zs,Drut:2010yn,Endres:2011er}.  In this letter we
consider finite-volume effects of composite particles in motion.  We discuss 
corrections to the binding energies of bound states in a moving frame.  We also show 
how the finite-volume scattering method is modified if one or both particles are 
composite.

We find topological phase corrections associated with the motion of bound states
in a periodic box.  These corrections have a universal dependence on momentum
determined by the number and mass of the constituents.  In asymptotically
large volumes the corrections are exponentially small and can be neglected.
However, it is often the case in large-scale lattice simulations that calculations 
are performed at volumes which are not asymptotically large.  Fortunately, we 
find that the corrections have a simple form which can be subtracted out from the 
analysis.

We start with another result derived by Lüscher~\cite{Luscher:1985dn}.  It
describes finite-volume corrections to the binding energy of two-body dimer states for
interactions with finite range.  For a dimer at rest, the shift in the energy
when placed in a periodic cube of volume $L^{3}$ is%
\begin{equation}
 \Delta E_{\vec{0}}(L)\approx\sum\limits_{|\vec{n}|=1}\int d^{3}r\,
 \phi_{\infty}^{\ast}(\vec{r})\,V(\vec{r})\,\phi_{\infty}(\vec{r}+\vec{n}L) \,.
\label{E2_0_L}%
\end{equation}
Here $V(\vec{r})$ is the interaction potential and $\phi_{\infty}$ is the
infinite-volume wavefunction as a function of the relative separation
$\vec{r}$.  The summation is over integer vectors $\vec{n}$ with magnitude $1$. 
Throughout our discussion, we assume that the energies and momenta are
non-relativistic.  For finite-range interactions Eq.~(\ref{E2_0_L}) gives a
correction which scales as $e^{-\kappa L}/L$ in the large volume limit, where
$\kappa$ is the binding momentum.

For general $N$-body bound states a straightforward generalization of Lüscher's
result yields%
\begin{align}
 \Delta E_{\vec{0}}(L) & \approx\sum\limits_{\sum_{j}
 \left\vert\vec{n}_{j}\right\vert=1}
 \int\left[{\displaystyle\prod\nolimits_{i}}d^{3}r_{i}\right]
 \phi_{\infty}^{\ast}(\vec{r}_{1},\cdots) \nonumber\\
 & \hspace{4em}\times\,V(\vec{r}_{1},\cdots)\,
 \phi_{\infty}(\vec{r}_{1}+\vec{n}_{1}L,\cdots) \,.
\label{EN_0_L}%
\end{align}
Here $\vec{r}_{i}$ are $N-1$ relative coordinates and $\vec{n}_{i}$ are
again integer vectors.  Although Eq.~(\ref{EN_0_L}) clearly does not apply to
relativistic quarks held by confinement within a single meson or baryon, these
corrections are useful for analyzing lattice calculations of hadronic molecules,
nuclei, and cold atomic bound states~\cite{Kreuzer:2008bi,Epelbaum:2009pd,
Kreuzer:2010ti,Epelbaum:2011md}.

We now consider a dimer moving in the same periodic cube with momentum
$2\pi\vec{k}/L$ for integer $\vec{k}$.  In the dimer wavefunction we can
factorize out the phase dependence due to the center-of-mass motion,%
\begin{equation}
 \psi_{L}\left(\vec{r}_{1},\vec{r}_{2}\right) = e^{i2\pi\alpha\vec{k}
 \cdot\vec{r}_{1}/L}e^{i2\pi(1-\alpha)\vec{k}\cdot\vec{r}_{2}/L}
 \phi_{L}(\vec{r}_{1}-\vec{r}_{2}) \,,
\label{full_2}%
\end{equation}
where $\alpha=m_{1}/(m_{1}+m_{2})$.  Since $\psi_{L}\left(\vec{r}_{1},
\vec{r}_{2}\right)$ is periodic in $\vec{r}_{1}$ and $\vec{r}_{2}$, $\phi_{L}$
gets a nontrivial phase for each winding around the toroidal topology of the
periodic cube,%
\begin{equation}
 \phi_{L}(\vec{r}+\vec{n}L)
 = e^{-i2\pi\alpha\vec{k}\cdot\vec{n}}\phi_{L}(\vec{r}) \,,
\label{rel_2}%
\end{equation}
for all integer $\vec{n}$.  We note that phase factors have been previously
discussed in connection with finite-volume scattering in moving
frames~\cite{Rummukainen:1995vs,Kim:2005gf,Feng:2011ah}.  However, the phase
factors have a qualitatively different effect on bound state wavefunctions which
simultaneously touch all wall boundaries.  Each phase twist induces a measurable
shift in the binding energy.  

When Eq.~(\ref{rel_2}) is combined with 
Eq.~(\ref{EN_0_L}), we find that the 
finite-volume correction is a sum of
sinusoidal functions of momentum.  For $S$-wave dimers with momentum
$2\pi\vec{k}/L$, the finite-volume correction has the form%
\begin{equation}
 \frac{\Delta E_{\vec{k}}(L)}{\Delta E_{\vec{0}}(L)}
 \approx \frac{1}{3}\sum_{l=1,2,3}\cos\left(2\pi\alpha k_{l}\right)
 \equiv \tau(\vec{k},\alpha) \,.
\label{E_tau_2}%
\end{equation}
Finite-volume corrections for higher angular momentum bound states at rest
have recently been discussed in~\cite{Konig:2011nz}.  From the results
presented there, it is straightforward to derive analogous results for
$\tau(\vec{k},\alpha)$ for dimers with angular momentum.

For bound states with more than two particles, the finite-volume correction has
the same general form.  For an $N$-body bound state with all equal masses and no
cluster substructure, the topological phase is%
\begin{equation}
 \phi_{L}(\cdots,\vec{r}_{i}+\vec{n}_{i}L,\cdots)
 = e^{-i2\pi\vec{k}\cdot\vec{n}_{i}/N}\phi_{L}(\cdots,\vec{r}_{i},\cdots) \,.
\label{rel_N}%
\end{equation}
For $S$-wave bound states we get%
\begin{equation}
 \frac{\Delta E_{\vec{k}}(L)}{\Delta E_{\vec{0}}(L)}
 \approx\tau\left(\vec{k},\frac{1}{N}\right) \,.
\label{E_tau_N}%
\end{equation}

For $N$-body bound states with a two-cluster substructure, one can apply the
two-body result, Eq.~(\ref{E_tau_2}), with $m_{1}$ and $m_{2}$ being the masses
of the two clusters.  For more complicated $N$-body bound states with more than
two clusters and/or particles with unequal masses, the same cosine functions as
in Eq.~(\ref{E_tau_2}) are also expected for $\Delta E_{\vec{k}}(L)/\Delta 
E_{\vec{0}}(L)$.  In these cases, however, more information is needed regarding
which particles or clusters occupy the tail of the bound state
wavefunction.  If this is unknown, then $\Delta E_{\vec{k}}(L)/\Delta
E_{\vec{0}}(L)$ can be extracted from numerical calculations, and an empirical
fit to cosine functions as in Eq.~(\ref{E_tau_2}) can yield structural information
about the tail of the bound state.

Our results presented above have already been adapted by 
Davoudi and Savage into a general method for reducing finite-volume 
errors for two-body bound states
such as the deuteron using boosted frames \cite{Davoudi:2011md}.  
The computational advantage of this approach is that finite-volume 
effects can be directly removed from lattice data without extrapolating to 
large lattice volumes.  
This is especially useful for the case with more than two constituents 
where the analytic form for the finite-volume $L$-dependence is \textit{a priori} unknown.

To illustrate the utility of the boosted-frame method, we consider lattice 
calculations of the triton at leading order in 
pionless effective field theory \cite{Bedaque:1999ve}.  
We use the leading-order lattice action defined in Eq.~(4.6) of 
Ref.~\cite{Lee:2008fa} with spatial lattice spacing $1.97~\text{fm}$ and temporal 
lattice spacing $1.32~\text{fm}/c$.  The two-body contact interactions, $C$ and 
$C_{I^2}$, are set to reproduce the physical neutron-proton scattering lengths, 
$a_{^{1}S_0} = -23.7~\text{fm}$ and $a_{^{3}S_1} = 5.4~\text{fm}$.  The 
three-body contact interaction, 
$D$, is determined by the triton energy at infinite volume, $-8.48~\text{MeV}$.

Using the Lanczos algorithm for sparse-matrix
eigenvector iteration~\cite{Lanczos:1950}, we have computed the triton energy as
a function of periodic cube length $L$ in spatial lattice units and 
momentum $2\pi\vec{k}/L$.  
In Table~\ref{triton} we show a comparison of lattice results for 
$\Delta E_{\vec{k}}(L)/\Delta E_{\vec{0}}(L)$
for the triton versus $\tau(\vec{k},\nicefrac13)$ for lattice sizes $L=6,7,8$.  
As seen in Table~\ref{triton}, the lattice
results approach $\tau(\vec{k},\nicefrac13)$ in the large-$L$ limit.
\begin{table}[tb]%
\begin{tabular}
[c]{|c|c|c|c|c|}\hline
$\vec{k}$ & $L=6$ & $L=7$ & $L=8$ & $\tau(\vec{k},1/3)$\\\hline\hline
$(1,0,0)$ & \multicolumn{1}{|r|}{$0.395$} & \multicolumn{1}{|r|}{$0.432$} &
\multicolumn{1}{|r|}{$0.458$} & \multicolumn{1}{|r|}{$0.500$}\\
$(1,1,0)$ & \multicolumn{1}{|r|}{$-0.035$} & \multicolumn{1}{|r|}{$-0.025$} &
\multicolumn{1}{|r|}{$-0.016$} & \multicolumn{1}{|r|}{$0.000$}\\
$(1,1,1)$ & \multicolumn{1}{|r|}{$-0.376$} & \multicolumn{1}{|r|}{$-0.413$} &
\multicolumn{1}{|r|}{$-0.442$} & \multicolumn{1}{|r|}{$-0.500$}\\\hline
\end{tabular}
\caption{Comparison of triton lattice results for $\Delta E_{\vec{k}}(L)/\Delta
E_{\vec{0}}(L)$ and $\tau(\vec{k},1/3)$ versus $L$.}%
\label{triton}%
\end{table}
We note that the leading finite-volume corrections vanish for $\vec{k}=(1,1,0)$.  
Therefore the calculation of the triton binding energy in this boosted frame should 
converge much more quickly in the limit of large $L$.  
In Table~\ref{triton2} we show the triton finite-volume 
energy corrections $\Delta E_{\vec{k}}(L)$ in MeV versus $L$ for $\vec{k}=(0,0,0)$ and 
$\vec{k}=(1,1,0)$.  We see that the finite-volume errors are reduced 
dramatically for $\vec{k}=(1,1,0)$.

\begin{table}[tb]%
\begin{tabular}
[c]{|c|c|c|c|c|}\hline
$\vec{k}$ & $L=5$ & $L=6$ & $L=7$ & $L=8$\\\hline\hline
$(0,0,0)$ & \multicolumn{1}{|r|}{$-0.603$} & \multicolumn{1}{|r|}{$-0.169$} &
\multicolumn{1}{|r|}{$-0.049$} & \multicolumn{1}{|r|}{$-0.015$}\\
$(1,1,0)$ & \multicolumn{1}{|r|}{$0.029$} & \multicolumn{1}{|r|}{$0.006$} &
\multicolumn{1}{|r|}{$0.001$} & \multicolumn{1}{|r|}{$0.0002$}\\\hline
\end{tabular}
\caption{Triton finite-volume 
energy corrections $\Delta E_{\vec{k}}(L)$ in MeV for $\vec
{k}=(0,0,0)$ and $\vec{k}=(1,1,0)$ versus $L$.}%
\label{triton2}
\end{table}

We now turn our attention to the scattering of composite states in a finite
periodic cube.  We consider the scattering between states $A$ and $B$ in the
center-of-mass frame.  Let $\mu_{AB}$ be the reduced mass, and let $E_{AB}(p,L)$
be the total energy of the $A-B$ scattering system with radial momentum $p$ in a
periodic cube of length $L$.  States $A$ and $B$ can be point particles or
composite bound states.  We assume that the constituent particles comprising the
states have finite range interactions.  The composite structures of $A$ and $B$,
however, will in general produce effective interactions with exponential tails
extending to infinity.

These tails generate exponentially-small finite-volume corrections to
$E_{AB}(p,L)$ associated with the binding energies of $A$ and $B$ separately
as well as the scattering of $A$ and $B$ together.  In this analysis, we focus
only on the exponential corrections to the binding energies.  This will be
useful in lattice simulations where the volume is not very large and the binding
energy shifts may be comparable to that of the scattering process being
measured.  We will not be concerned with exponentially small corrections to the
scattering of $A$ and $B$.  If the interactions between $A$ and $B$ are very
strong, then it is theoretically possible that the finite-volume scattering
corrections we neglect are comparable to the binding energy shifts.  However, in
such cases the part of the energy shift due to scattering which is not
exponentially suppressed will be much larger still, and so the loss of accuracy
in the scattering analysis will be small.

In order to calculate finite-volume corrections due to the binding energy, it
suffices to consider singular solutions of the free Helmholtz equation.  Let
$\vec{r}$ be the separation between the center of masses of the two states.  In
the following we assume that $p$ is sufficiently small so that angular momentum
mixing with higher-order singular solutions can be neglected.  For $S$-wave
scattering between states $A$ and $B$ with radial momentum $p$, the
position-space scattering wavefunction is%
\begin{equation}
 \left\langle \vec{r}\right.\left\vert \Psi_{p}\right\rangle
 = c\,{\displaystyle\sum_{\vec{k}}}\frac{e^{i\frac{2\pi\vec{k}}{L}\cdot\vec{r}}}
 {\big(2\pi\vec{k}/L\big)^{2}-p^{2}}
\label{scatteringwave}%
\end{equation}
with some normalization constant $c$.

We let $E_{\vec{k}}^{A}(L)$ and $E_{-\vec{k}}^{B}(L)$ be the finite-volume
energies due to binding for bound states $A$ and $B$ with momenta $2\pi\vec
{k}/L$ and $-2\pi\vec{k}/L$, respectively.  For point particles without internal
structure, these energies are by definition zero for all momenta.  The total
energy $E_{AB}(p,L)$ is then%
\begin{equation}
 E_{AB}(p,L) = \frac{\left\langle\Psi_{p}\right\vert H
 \left\vert\Psi_{p}\right\rangle }{\left\langle \Psi_{p}\right.
 \left\vert \Psi_{p}\right\rangle}
 = \frac{1}{\mathcal N}  
 \sum\limits_{\vec{k}}\tfrac{\frac{p^{2}}{2\mu_{AB}}+E_{\vec{k}}^{A}(L)
 +E_{-\vec{k}}^{B}(L)}{\left(\vec{k}^{2}-\eta\right)^{2}} \,,
\label{scatt_1}%
\end{equation}
where ${\mathcal N} = \sum_{\vec{k}}\big(\vec{k}^{2}-\eta\big)^{-2}$
and $\eta=p^{2}L^{2}/(2\pi)^{2}$.  The finite-volume correction can be written
as%
\begin{multline}
 E_{AB}(p,L)-E_{AB}(p,\infty) \\
 = \tau_{A}(\eta)\,\Delta E_{\vec{0}}^{A}(L)
 + \tau_{B}(\eta)\,\Delta E_{\vec{0}}^{B}(L) \,,
\label{scatt_2}%
\end{multline}
where $\Delta E_{\vec{0}}^{A}(L)$ and $\Delta E_{\vec{0}}^{B}(L)$ are the
finite-volume corrections for states $A$ and $B$ at rest, and we have defined
the topological volume factor%
\begin{equation}
\tau(\eta)=\frac{1}{\mathcal{N}}{\sum_{\vec{k}}}\tfrac{\sum_{l=1,2,3}
\cos\left(  2\pi\alpha k_{l}\right)  }{3\left(  \vec{k}^{2}-\eta\right)^{2}}\,.
\end{equation}
The analysis can be generalized to higher angular momentum scattering states
using an extension of Lüscher's scattering relation to higher orbital angular
momentum~\cite{Bernard:2008ax,Luu:2011ep}.

The finite-volume correction in Eq.~(\ref{scatt_2}) has nothing to do with the
interaction between states $A$ and $B$ and should therefore be subtracted from
the total energy before using Lüscher's scattering relation.  This subtraction
should reduce systematic errors in lattice calculations. 

As an example to test the method, we consider fermion-dimer scattering for 
two-component fermions.  The physical process which we study corresponds with 
spin-quartet scattering between a neutron and deuteron.  In that case the 
two fermion components should be regarded as protons and neutrons with the same 
spin.

We use the same lattice Hamiltonians as in Ref.~\cite{Bour:2011xt}, except in 
that case the scattering length was tuned to infinity.  As in Ref.~\cite{Bour:2011xt}, 
we consider
two different lattice Hamiltonians, each of which produces a shallow dimer in
the continuum limit.  The first Hamiltonian, $H_1$, contains only a local contact 
interaction between fermions.  The second Hamiltonian, $H_2$, contains a contact 
interaction as well as nearest-neighbor interactions.  These are used 
to tune the binding energy of the dimer while also setting the effective 
range parameter to zero.  Both lattice Hamiltonians reproduce the same 
continuum limit of fermions with attractive zero-range interactions.

We now focus on the fermion-dimer system.  This corresponds with a 
neutron together with a deuteron in the spin-quartet 
channel in pionless effective field theory at leading order.
Experimental 
measurements find a quartet scattering length $^{4}a_{nd}=6.35(2)~\text{fm}$
\cite{Dilg:1971}.  This can be expressed as a fraction of the spin-triplet
neutron-proton scattering length, $^{4}a_{nd}/^{3}a_{np}=1.17(1)$.
A more detailed calculation including 
interaction range effects obtains $^{4}a_{nd}=6.33(10)~\text{fm}$ 
\cite{BvK:1998,BHvK:1998}, in full agreement with experimental values.

We calculate fermion-dimer scattering on the lattice using Lüscher's
finite-volume formula.  Using the Lanczos algorithm we have computed the ground
state energy for the fermion-dimer system on periodic cubes for
six different lattice spacings $a_{\text{latt}}$.  For each lattice spacing we
consider periodic volumes $L^{3}$ ranging from $L=6$ to $L=17$ lattice units.  
From these energies we determine the low-energy parameters for
fermion-dimer scattering and extrapolate to the continuum limit.  The full details
of this calculation will be described in a forthcoming publication, and we just
summarize the results here.

\begin{figure}[ptb]
\centering\includegraphics[width=0.45\textwidth]{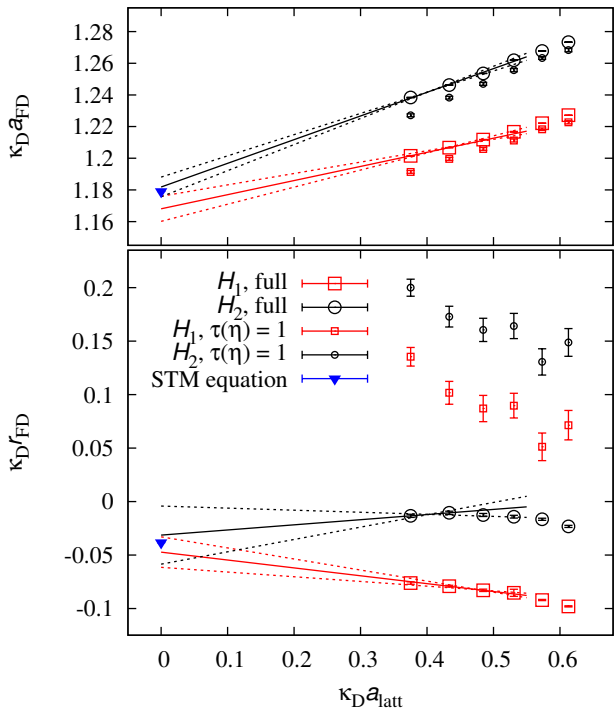}
\caption{(Color online) Lattice results and continuum extrapolation
with error estimates for the fermion-dimer scattering length (top) and effective
range parameter (bottom). For comparison we show the continuum results obtained
via the Skorniakov-Ter-Martirosian equation.}%
\label{fig:a_r}
\end{figure}

Results for the fermion-dimer scattering length, $a_{\text{FD}}$, and the effective range
parameter, $r_{\text{FD}}$, are shown in Fig.~\ref{fig:a_r}.  In each case, we
extrapolate to the continuum limit and write final results as dimensionless
combinations multiplied by powers of the dimer binding momentum,
$\kappa_{\text{D}}$.  In the shallow binding limit $\kappa_{\text{D}}$ equals 
the reciprocal of the fermion-fermion scattering length.  To see the effect of 
the topological volume correction we
have done the full calculation using the correct topological factor
$\tau(\eta)$, as well as a faulty calculation which replaces $\tau(\eta)$ by
$1$. For comparison, we show the continuum results, $a_{\text{FD}} \kappa_{\text{D}} =
1.17907(1)$ and $r_{\text{FD}} \kappa_{\text{D}} = -0.0383(3)$, obtained via the
Skorniakov-Ter-Martirosian (STM) integral
equation~\cite{Skorniakov:1957,Braaten:2004rn}.  We find that the small size of
the effective range parameter requires a fit to low-energy scattering that
includes the shape parameter, which was not done in earlier calculations of the
effective range parameter~\cite{vonStecher:2008a}.  We note also the agreement 
with the experimental value $^{4}a_{nd}/^{3}a_{np}=1.17(1)$.

The results in
Fig.~\ref{fig:a_r} show that the inclusion of the
topological volume factor $\tau(\eta)$ is essential for obtaining the correct
continuum limit.  In all cases the continuum extrapolations for $H_{1}$ and
$H_{2}$ agree within error bars.  However, the correct $\tau(\eta)$ factor is
needed to reproduce the STM equation result.  The correction is small for the
scattering length, but quite large for the effective range parameter.

We expect the topological volume factor to have important effects in other
lattice calculations of scattering for composite bound states.  The analysis
presented here should provide a simple but general method for improving the
accuracy of bound-state scattering calculations.  The list of possible
applications is quite extensive and include lattice calculations involving the
scattering of nucleons upon nuclei, the scattering of nuclei, Compton
scattering and electroweak probes upon nuclei, mesonic and baryonic scattering
upon hadronic molecules, and few-body scattering in cold atomic systems.
An application of this method to dimer-dimer scattering is in progress.\medskip

\begin{acknowledgments}
We thank Doerte Blume and Martin Savage for useful discussions.  Partial
financial support from the Deutsche Forschungsgemeinschaft (SFB/TR 16),
Helmholtz Association (contract number VH-VI-231), BMBF~(grant 06BN9006), and
U.S. Department of Energy (DE-FG02-03ER41260) are acknowledged.  This work
was further supported by the EU HadronPhysics2 project ``Study of strongly
interacting matter'', and S.K. was supported by the ``Studien\-stiftung des
deutschen Volkes'' and by the Bonn-Cologne Graduate School of Physics and
Astronomy.
\end{acknowledgments}

\end{document}